# Fluid Antennas: Reshaping Intrinsic Properties for Flexible Radiation Characteristics in Intelligent Wireless Networks

Wen-Jun Lu, *Senior Member, IEEE,* Chun-Xing He, *Graduate Student Member, IEEE,* Yongxu Zhu, *Senior Member, IEEE*, Kin-Fai Tong, *Fellow, IEEE*, Kai-Kit Wong, *Fellow, IEEE*, Hyundong Shin, *Fellow, IEEE*, and Tie Jun Cui, *Fellow, IEEE*

*Abstract* — Fluid antennas present a relatively new idea for harnessing the fading and interference issues in multiple user wireless systems, such as 6G. Here, we systematically compare their unique radiation beam forming mechanism to the existing multiple-antenna systems in a wireless system. Subsequently, a unified mathematical model for fluid antennas is deduced based on the eigenmode theory. As mathematically derived from the multimode resonant theory, the spectral expansion model of any antennas which occupy variable spaces and have changeable feeding schemes can be generalized as 'fluid antennas'. Non-liquid and liquid fluid antenna examples are presented, simulated and discussed. The symmetry or modal parity of eigenmodes is explored as an additional degree of freedom to design the fluid antennas for future wireless systems. As conceptually deduced and illustrated, the multi-dimensional and continuously adaptive ability of eigenmodes can be considered as the most fundamental intrinsic characteristic of the fluid antenna systems. It opens an uncharted area in the developments of intelligent antennas (IAs), which brings more flexibility to on-demand antenna beam/null manipulating techniques for future wireless applications.

## I. Introduction

Multiple-input-multiple-output (MIMO) [1] and massive MIMO (mMIMO) [2] lay the foundation of modern high speed and large capacity wireless communications. In order to achieve more flexibly agile beam and higher multiplexing gain for high speed and large capacity transmissions and space-division multiple access, a concept of reconfigurable intelligent surfaces (RISs) [3, 4] is recently considered as an emerging technology in future wireless communications. Basically, RISs are classified into two different types according to the size of meta-atom. In non-resonant case, meta-atoms with electrically small size of less than $\lambda/2\pi$ (where $\lambda$ denotes the carrier wavelength) are arranged to create a reactive surface [5]. Half-wavelength radiating elements in the resonant case can yield spatially fed phased array [5]. To some extent, RISs share a similar origin to mMIMO and the time-modulated arrays (TMAs) [6]. The beam agility for RISs, mMIMO, and TMAs with fixed layouts can be commonly described by the principle of pattern multiplication in discrete antenna arrays, as shown in Fig. 1. When the number of elements $N$ approaches infinity and the inter-element separation $d$ approaches 0, a discrete antenna array evolves into a continuous one, known as a continuous-aperture MIMO. If a finite but continuous array becomes repositionable, i.e., near-continuously switching the radiator's position in a predefined area to a position where the channel conditions are desirable, it naturally gives rise to the fluid antenna (FA).

The FA concept enhances the spatial diversity and can be translated into interference reduction by exploiting the fading opportunity in multiuser communications. Ideally, the occupied space, the boundary conditions and feed schemes of FAs can be entirely flexible. This implies that FAs could be reshaped as desired in theory, with its intrinsic resonant properties such as resonant frequency, resonant mode, and radiation pattern could be arbitrarily adjusted.

In Fig, 1, we compare FA with RIS and mMIMO/TMA based on their respective mechanisms of beam reconfiguration. Here, all discrete elements in RISs, mMIMO and TMAs occupy specific positions with fixed inter-element separations of $D_x$, $D_y$, and $D_z$, which may respectively yield locked spatial phase shift terms of $kD_x\sin\theta\cos\varphi$, $kD_y\sin\theta\sin\varphi$, and $kD_z\cos\theta$ in the array factor (AF), in which $k$ represents the operational wave number, $\theta$ and $\varphi$ are, respectively, the elevation and azimuth angles,

Manuscript received on Sep. 15, 2024. This work was partly supported by National Key Research and Development Program of China under grant no. 2021YFE0205900, Key Technologies R&D Program of Jiangsu (Prospective and Key Technologies for Industry) (BE2022067 and BE2022067-2), and Natural Science Foundation of China under grant no. 61871233, the work of K. K. Wong and K. F. Tong is supported by the Engineering and Physical Sciences Research Council (EPSRC) under grant EP/W026813/1.

W. -J. Lu and C. -X. He are with the Jiangsu Key Laboratory of Wireless Communications, Nanjing University of Posts and Telecommunications, Nanjing 210003, China (Corresponding author: *Wen-Jun Lu*, e-mail: wjlu@njupt.edu.cn, *Kai-Kit Wong*, kai-kit.wong@ucl.ac.uk). Y. Zhu is with the National Key Laboratory of Mobile Communications, Southeast University, Nanjing 211111, China (e-mail: yongxu.zhu@seu.edu.cn). T. J. Cui is with the State Key Laboratory of Millimeter Waves, Southeast University, Nanjing 211111, China (e-mail: tjcui@seu.edu.cn). K. -F. Tong and K. -K Wong are with the Department of Electronic and Electrical Engineering, University College London, London WC1E 7JE, U.K. (e-mail: {k.tong,kai-kit.wong}@ucl.ac.uk), K. K. Wong and Hyundong Shin are affiliated with the Department of Electronic Engineering, Kyung Hee University, Yongin-si, Gyeonggi-do 17104, Korea.(e-mail: hshin@khu.ac.kr)

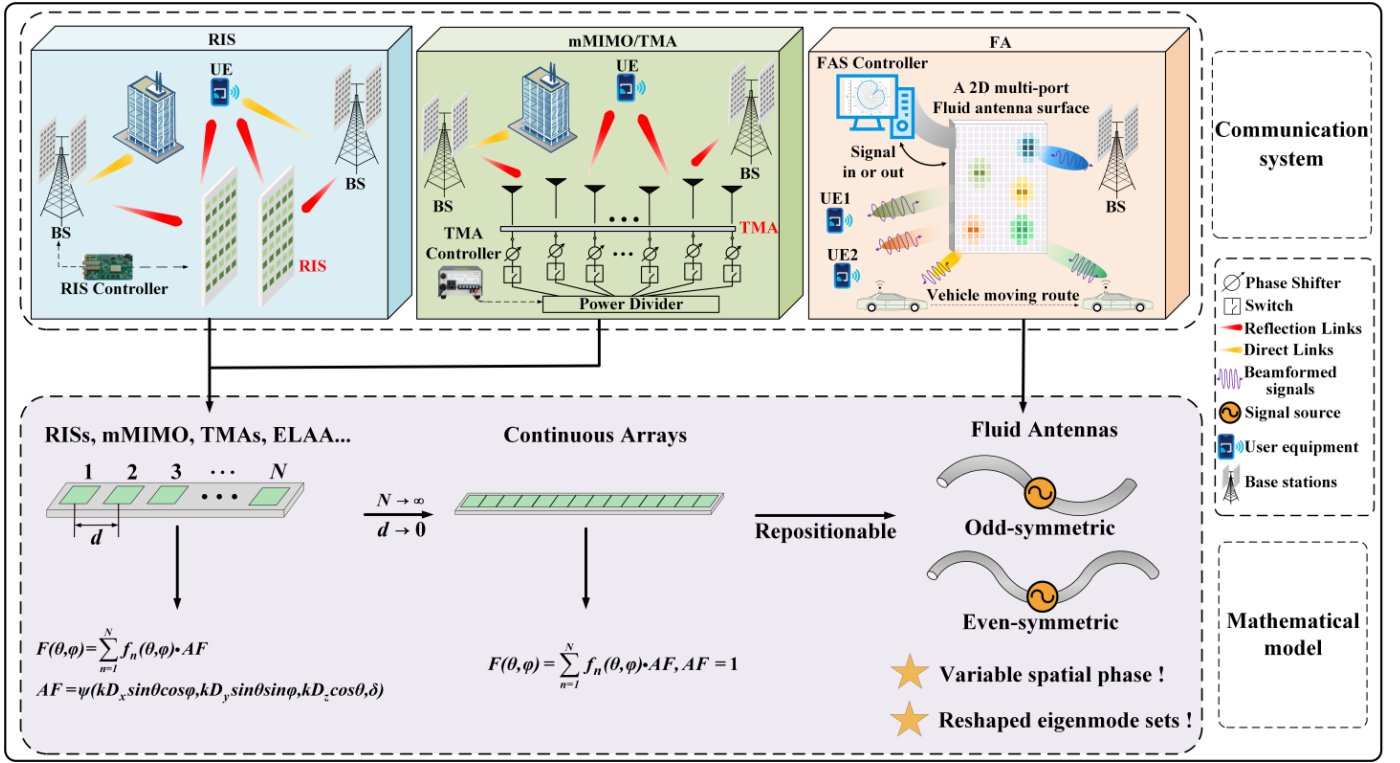

Fig. 1: Graphical illustration of RIS, mMIMO/TMA, FA and their corresponding mathematical model.

$f_n(\theta, \varphi)$ is the radiation pattern of the $n$th radiator, and $\Psi(\cdot)$ is the general expression of phase shift function. Electrically tunable elements loaded to a fixed-layout array aim at varying the temporal phase shift term $\delta$ in $AF$, while they rarely change the intrinsic resonant modes of a radiator or the array, but just affect the excitation and combination of these modes.

Unlike RIS and mMIMO/TMA with discrete and fixed layouts, the flexibility in shape of FA brings an agile radiation pattern $F(\theta, \varphi)$ in the sense of a continuous array. For a single FA, $AF$ is unity (i.e., $AF=1$), and the radiation pattern $F(\theta, \varphi)$ is formed by the superposition of eigen-radiation patterns of $f_n(\theta, \varphi)$, generated by multiple excited resonant eigenmodes. In other words, the beam reconfigurability is enabled and governed by the intrinsic property of FA. This concept has inspired new beam forming techniques for user equipment (UE) with simple antenna deployment in limited space by deploying non-solid radiating antennas [7]. Recent studies have demonstrated that the FA systems can provide an agile beam or null to UE by shifting the position of fluid radiators [8, 9]. When using FA in multiple access environment, i.e., fluid antenna multiple access (FAMA), significant interference elimination and multiplexing gain enhancement are observed [9]. Another key member of FA is pixel reconfigurable antenna, which can realize adjustable intrinsic properties [10] to produce variable wave numbers of $k_v$. Additionally, movable antennas [11] follow the conventional approach to tune the spatial phase shifts of the discrete antenna array by replacing the fixed inter-element separations $D_x$, $D_y$, and $D_z$ with variable $D_{xv}$, $D_{yv}$, and $D_{zv}$. In the preceding analyses on agile beam mechanism of RIS, mMIMO/TMA and movable antennas [3, 4, 6, 10, 11], the radiator is assumed to be single-mode resonant, with fixed and less frequency-dispersive radiation pattern $f(\theta, \varphi)$, while FA is multimode resonant. In the next section, we will further explore the mechanism of continuous beam reconfigurability in FA using the eigen-mode theory and multimode resonant antenna model.

## II. UNIFIED MATHEMATICAL MODEL FOR FA

As illustrated in Fig. 1, the radiation pattern of a FA is co-dominated by multiple simultaneously excited resonant eigenmodes. Basically, flexibly variable boundary conditions (BCs) of FA yield different eigenmode sets, as well as feed schemes, and thus they lead to reshapable radiation behavior. FA can be distinguished from RIS [3-4], mMIMO [2] or TMA [6], and movable antennas [11] by the agility in the eigen-spectral domain. Based on the eigenmode theory [5], a unified mathematical model of FAs can be deduced as

$$(\nabla^2 + k_{nmp}^2)I(x,y,z;x',y',z') = \delta(x-x')\delta(y-y')\delta(z-z')$$
where (1a)
$$x, x' \in [-L, L], y, y' \in [-W, W], z, z' \in [-H, H],$$
and
$$aI + b\frac{\partial I}{\partial r}\Big|_{\partial\Omega} = 0 \quad (r = x, y, z)$$

$$I(x,y,z;x',y',z') = \sum_{n=1}^{\infty}\sum_{m=0}^{\infty}\sum_{p=0}^{\infty}\frac{I_{nmp}(x',y',z')I_{nmp}(x,y,z)}{k^2 - k_{nmp}^2}$$
where (1b)
$$I_{nmp}(x,y,z) = \frac{\cos}{\sin}\frac{n\pi}{2L}x\frac{\cos}{\sin}\frac{m\pi}{2W}y\frac{\cos}{\sin}\frac{p\pi}{2H}z,$$
and
$$k_{nmp}^2 = \sqrt{(\frac{n\pi}{2L})^2 + (\frac{m\pi}{2W})^2 + (\frac{p\pi}{2H})^2}, n=1,2,\ldots \quad m,p=0,1,2,\ldots$$

The eigenmodes of a resonant antenna can be computed, identified and used by solving the homogeneous wave equation under closed BCs in a manner of solving interior waveguide problems [5, 12-13]. For FAs, since the fluid radiator position and associated BCs are variable, the eigenmode sets have to be dynamically changed and variable feed schemes have to be employed to derive the eigenmode combinations under the multimode resonance [12- 13]. To illustrate this idea, a simple mathematical model for describing the eigenmodes of FAs with variable shapes and positions is deduced in the rectangular coordinate system here.

The eigen-current (electric or magnetic) modes of a rectangular radiator can be attained by solving the partial differential equation with homogeneous BCs at the scalar form, as expressed by (1a). A FA is described as a radiator fed at ($x'$, $y'$, $z'$) by an external port, which can be emulated by a Dirac impulse function, which is even-symmetric about the feed coordinate. Theoretically, the excitation function with different symmetries can be arbitrarily set to emulate different feed schemes [13]. Unlike rigid antennas with fixed shapes, the radiator under investigation occupies a finite space $\Omega$ but variable length of $2L$, width of $2W$ and height of $2H$ in $x$-, $y$-, and $z$-directions, respectively, or it should have at least one flexibly variable dimension. On its boundary $\partial\Omega$, the BCs can be expressed by the value of the eigenmode function and/or its first-order directional derivative, where $a$ and $b$ are constant coefficients. For different $a$ and $b$, it yields the first kind (Dirichlet, $a=1$ and $b=0$), the second kind (Neumann, $a=0$ and $b=1$), and the third kind (Robin, $a$, $b \neq 0$) BC problems respectively. The eigenmodes are independent on the excitation but rely on the radiator's shape and BC only. In the rectangular coordinate system, the homogeneous wave equation in (1a) with associated homogeneous BCs can be solved by using the Fourier series expansion method in the spectral domain, such that the surface current distribution $I(x, y, z, x', y', z')$ is expressed by the spectral expansion form in terms of harmonic eigenmodes $I_{nmp}(x, y, z)$ ($n=1, 2, 3..., m, p=0, 1, 2, 3,...$) in (1b). Once the feed scheme is known, the amplitude of each excited eigenmode can be computed to determine the total electric surface current distribution. Thus, the radiation patterns of each eigenmode, as well as the combined total far-field pattern $f(\theta, \varphi)$ can be transformed from the electric/magnetic current eigenmodes [12, 13] and then superposed accordingly.

The mathematical and physical meaning of (1) is intuitive and manifest: Once the position of the fluid radiator of a FA varies (at least, in one dimension), the eigenmode family $\{I_{nmp}(x, y, z)\}$ subsequently varies, as well as the port and the feed position ($x'$, $y'$, $z'$) accordingly varies. Therefore, distinct radiation behavior can be dynamically manipulated by the different combinations of eigenmode functions. This is exactly the philosophy of 'Be Water' [8, 11]: Intuitively, a FA can be arbitrarily reshaped. Mathematically, a FA can reconfigure its eigenmode family $\{I_{nmp}(x, y, z)\}$, the excitation function and its location ($x'$, $y'$, $z'$), such that the excitation status of eigenmodes can be adjusted to produce distinct radiation behaviors as desired, just like the 'unity AF case' described in Fig. 1. Figs. 2(a) and 2(b) conceptually illustrate the concept:

The simple straight fluid metal, center-fed, half-wavelength dipole under differential mode excitation can be reshaped into a full-wavelength one under dual-port, common mode excitation. In this manner, the doughnut-shaped radiation pattern in Fig. 2(a) can be reshaped into the four-lobe one in Fig. 2(b). As such, the mathematical essence of FAs can be briefly described by (1), with its physical insights intuitively illustrated by Fig. 2.

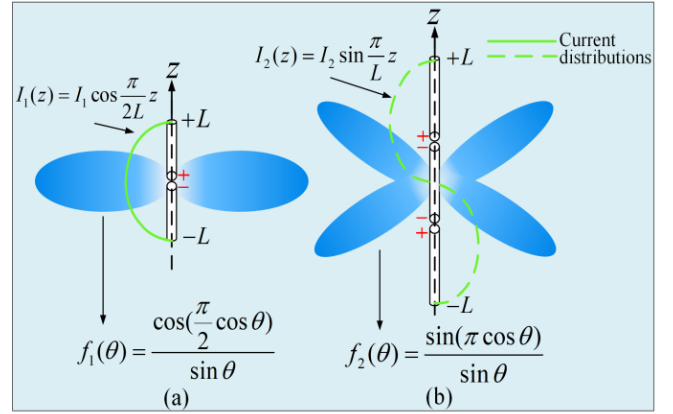

Fig. 2 Conceptual fluid linear dipole with traversable ports, corresponding to eigenmode functions and radiation patterns: (a) even-symmetric mode, and (b) odd-symmetric mode.

The unified mathematical model based on eigenmode theory reveals the ultimate difference between FAs, mMIMO/TMAs, and RISs. With reference to (1), we may call phased arrays (active, circuital fed cases of TMA and mMIMO, and passive, spatially fed case of RIS) that with their temporal phase shift term tuned to yield steerable beams, commonly known as '*smart antennas*'. Similarly, we are also able to name the FAs that rely on their reshapable intrinsic properties in multiple dimensions (i.e., eigenmode family, BCs, feed schemes, etc.) to yield finely controllable beam/null as '*intelligent antennas*'. Under the guidance of semi-analytical, mode synthesis antenna design approaches [12, 13], novel FAs under multimode resonance can be designed by properly manipulating their eigenmode sets and feed schemes in the future.

III. GENERALIZATION OF FA

The term 'FA' can be generalized as accordingly expected: Antennas that occupy variable space, with their eigenmode family and feed scheme flexibly modified accordingly, can demonstrate the *Be Water* philosophy, even if it is not fabricated by liquid substance like liquid alloy, salted water or other conductive and dielectric liquids. The term 'fluid' is not limited to 'liquid' [7] or 'movable' [11] anymore, but is defined mathematically by the dynamically tunable eigenmodes, i.e., the intrinsic resonant properties of itself. To some extent, any antennas that can be reshapable, have their forms to reconfigure their eigenmode families, can be reasonably classified as 'FAs', regardless of their fabricated materials. According to the unified mathematical model, FAs include, but are not limited to, several possible types, as discussed below.

i) The first type is 'plasma FA', which can evolve from traditional plasma antennas [14]. On one hand, plasma antennas

can be realized by plasma discharge tubes or fluorescent lamps, such that they can be combined to yield different radiation behaviors. On the other hand, fluorescent lamps can be combined with other types of fixed antennas to yield all kinds of beam agile antennas [15].

ii) The second type is 'thermal sensitive antennas' or 'solid/liquid state FAs', which may be fabricated by low-melting materials. For example, we can use a low-melting point alloy, such as mercury, bismuth and galinstan, to design a solid antenna operating at temperature $T_0$ and resonant frequency $f_0$. When the antenna is being heated to a certain temperature $T_1$, the alloy may melt to produce a liquid metal antenna detuned at frequency $f_1$. By detecting the detuned frequency shift of the antenna, ambient temperature variation can be sensed. Besides alloy, we can also use non-crystal, low-melting point dielectric materials like wax, rosin, asphalt, or even glass to design dielectric resonator antennas (DRAs) having thermal sensing ability. In these ways, ambient temperature variation can be sensed by detecting the detuned frequency shift or varied radiation behavior of these antennas. Such conceptual thermal sensitive antennas can be classified as the solid/liquid state FAs.

iii) The third type is hybrid-state FAs. They can be treated as the combinations of FAs with different states. Multiple FAs having different states can be combined as desired to yield more novel FAs with distinctive functionalities.

properly tuned to different combinations, distinctive unidirectional patterns can be attained, as denoted by the dashed and dotted lines in Fig. 3(c). Therefore, the prototype intuitively presents the conceptual design of plasma FAs with electronically switchable omnidirectional and unidirectional beams. In this case, the port is stationary, rather than movable, with the usable mode switched by excited plasma cylinders. Compared to liquid state FAs [9], the plasma FA may be free from the obstacle of liquid inertia [8, 9], since the on-off states of lamps can be switched electronically and encoded as desired to produce the distinct operational modes. The downsides are fragility and relatively bulky in size. Hence, plasma FAs would be considered a better candidate for fixed wireless access points than for compact, mobile terminals.

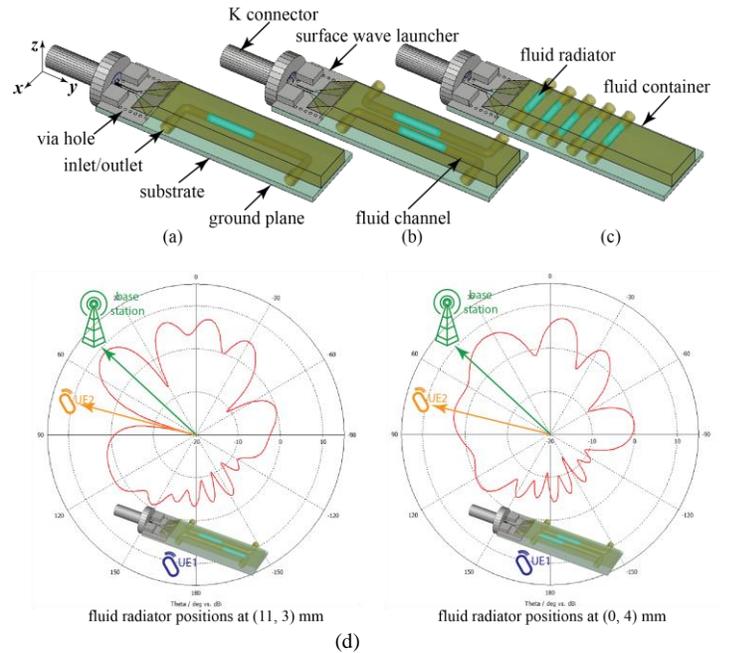

(d)
Fig. 4 Surface wave enabled liquid metal antenna at mmWave band for dynamical null manipulation: (a) basic configuration, (b) dual-channel coupled configuration, (c) multiple channel coupled, and (d) beam configuration via repositioning the liquid metal cylinders.

Fig. 4 illustrates a surface wave enabled FA system at 24-30 GHz. The liquid metal in the fluid container is excited by a printed circuit board (PCB) surface wave launcher. The radiation null is dynamically positioned by varying the shifted position $L_{shift}$ of the liquid metal to yield interference mitigation from other UE. On the other hand, the liquid metal radiator can be shifted to a different position when a desired UE approaches the null direction. When $L_{shift}$ continuously varies, it yields smoothly controllable beam/null directions functionality, which is beneficial for anti-fading and interference mitigation in mobile communications.

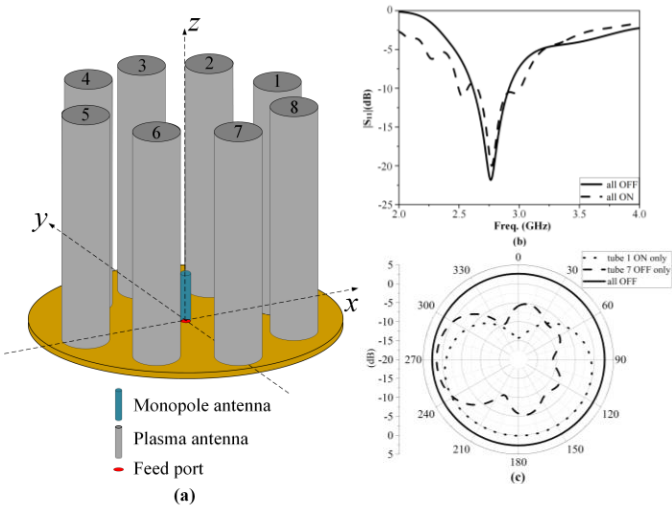

Fig. 3 Simulated plasma FA with electrically tunable omnidirectional and unidirectional radiation patterns: (a) configuration, (b) reflection coefficients, and (c) azimuth radiation patterns.

A preliminary example of plasma FA is simulated to demonstrate its radiation beam forming ability [15]. As shown in Fig. 3(a), the plasma FA is composed of eight, cylindrical fluorescent lamps and a centered monopole. When all fluorescent lamps are on to produce a plasma FA, the resonant band is slightly expanded at the centered frequency, as illustrated in Fig. 3(b). When all fluorescent lamps are off, the antenna exhibits a typical omnidirectional azimuth pattern just like traditional monopole antennas, as indicated by the solid line in Fig. 3(c). When the on-off states of the lamps are

In this section, two conceptually simulated prototypes (i.e., one non-liquid and one liquid) FAs are raised and discussed herein. However, evidently, FAs should never limit themselves to the aforementioned categories. Motivated by different applications in the future Internet-of-Everything (IoE), there will be many other possible FAs that are more suitable to certain applications by incorporating more advanced materials and fabrication techniques. For example, the pixel reconfigurable antenna approach in [10] is particularly attractive when switching delay is a major factor.

## IV. FUTURE TRENDS AND DISCUSSIONS

Latest advancement of multimode resonant antenna designs yields the 'generalized odd-even mode theory' [12, 13], where the symmetry of eigenmodes is recognized as an important factor in antenna designs. In the theory, eigenmodes can be classified into even-symmetric and odd-symmetric sets. Also, the excitation function emulating the feed line can be classified in the same manner. Hence the even-symmetric, Dirac impulse function in (1a) can be replaced by its odd-symmetric first-order derivative of the Dirac doublet function, or by their linear combinations [13]. As illustrated in Fig. 2, the symmetry or parity of eigenmodes is an important factor which dominates the radiation behavior of antenna systems. Mathematically, when an antenna is fed asymmetrically, the excitation function will be partially projected on the modal mismatched resonant modes of the antenna. Such effect can be expressed by the cross-coupled term in the interior Green's function [13], i.e., the surface current distribution given by (1b) in the spectral expansion form. Physically, when modal mismatching between feed line and antenna exists, unbalanced currents may be excited on the outer surface of the feed line. Thus, it leads to unbalance phenomenon in antenna systems [13]. In future FA designs, unbalanced currents may be utilized as new degree of freedoms in design. In terms of modal parity and modal mismatching in antenna systems, possible trends for novel FAs are briefly discussed below.

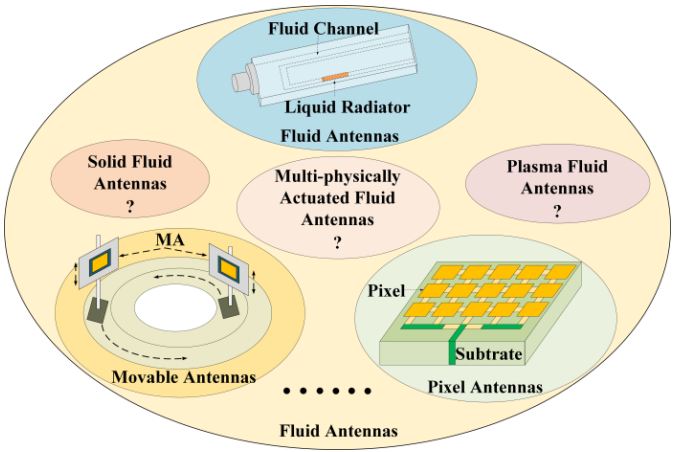

Fig. 5: Generalization of the FA family.

i) **FA with variable modal parity**—Parity of eigenmodes is one of the most important intrinsic properties of FAs. One FA may exhibit different modal parities in distinct dimensions. Therefore, the radiation behavior of FAs can be designed more systematically. By adjusting the BC in each dimension, the symmetry of eigenmodes can be interchanged and rearranged. By properly setting multiple ports with different symmetries, modal parity in each dimension can be excited and combined accordingly. In this manner, FAs with variable modal parity and distinct radiation patterns can be designed.

ii) **Unbalanced FA**—When the modal parity mismatching exists in a FA, unbalanced currents will occur in the antenna system but it remains an open problem [13]. The transformation of unbalanced currents into positive is useful to create new degrees of freedom for antenna design. The prompt creation of on-demand 'currents' which can dynamically adjust the radiation patterns has a high potential. Self-balanced [12, 13] techniques to FAs would be a challenging problem.

iii) **FA systems**—In future IoE, different kinds of FA techniques can be combined to produce different antenna systems. The design of FA systems should be inspired and boosted by diverse access schemes, networking strategies, and applications in IoE. Plasma, liquid and solid state FAs can be combined as needed to construct novel wireless systems with different functions, i.e., sensing, multiple access, networking, etc., in terms of their eigenmodes, excitation schemes, and modal parities.

Fig. 5 illustrates the generalization of FA family. It covers existing liquid FAs [8, 9], pixel antennas [10], and movable antennas [11], and prospective plasma and solid state FAs. As expected, many possible new members of FAs can be designed and implemented in the future to extend the framework of the FA family. New materials (e.g., phase change materials, flexible materials, etc.) and advanced fabrication process (e.g., 3-D printing) enable more novel FAs with smoothly reshaped eigenmodes. Novel composite materials also bring new radiation mechanism to the FAs under multi-physically (e.g., acoustically, mechanically, thermally, etc.) actuated resonances. In that way, FAs could be modeled and manipulated by simultaneously solving the coupled acoustical, mechanical, and thermal equations and the Maxwell's equations. Therefore, more research for FA development is required, and explored by solving the complex multi-physics problems.

TABLE I
COMPARISON OF FA, RIS, AND MMIMO/TMA

|  | FA | RIS | mMIMO/TMA |
| --- | --- | --- | --- |
| Shape/layout | Variable | Fixed | Fixed |
| Radiation pattern forming mechanism | Smoothly repositionable eigenmodes, feeds and modal parity | AF | AF |
| State of matter | Liquid, solid, plasma and hybrid | Solid | Solid |

Table I summarizes the main differences between FAs, RISs and mMIMO/TMA in terms of shape, radiation pattern forming mechanism and state of matter. It is clearly seen that FA should be distinctive in all aspects from its counterparts. In future FA system designs, on-demand and smooth beam/null steering techniques would be the most challenging and comprehensive problem that covers multiple disciplines including applied mathematics, applied physics, electromagnetics, antennas and propagation, wireless communication, new materials, advanced fabrication, and more.

## V. Conclusion

The mathematical essence and physical insight of FAs are anatomized in the perspective of eigenmode theory and modal parity, i.e., the intrinsic properties of FAs. As revealed by a unified mathematical model of FAs, the beam agile mechanism of FAs has been explained and compared to those of RISs and TMAs. It is then understandable that FAs exhibit different radiation beam forming principles to RISs and TMAs in the perspective of eigenmode theory and modal parity. Fundamentally, the term 'FA' [8, 9] is not limited to 'liquid antennas' [7] or 'movable antennas' [11]. Any antennas with dynamically controllable BCs, variable occupied space (or topological structure), feed scheme and eigenmode parity, can be classified as 'FAs', regardless of their fabricated materials. Thus, FA is generalized as a collective concept that describes 'antennas with repositionable radiation behaviors'. The multi-dimensional agility in eigenmode functions, modal parity and feed positions are the most crucial characteristics of FA. Under this generalized framework, the *smoothly reshaped eigenmodes* should be the intrinsic property of FAs.

**Biography**

**Wen-Jun Lu** is a Professor in the Jiangsu Key Laboratory of Wireless Communications, Nanjing University of Posts and Telecommunications (NUPT). His research interests are antenna theory and channel modeling techniques in wireless communications.

**Chun-Xing He** is a postgraduate student in the Jiangsu Key Laboratory of Wireless Communications, Nanjing University of Posts and Telecommunications (NUPT). His research interests are antenna theory and channel modeling techniques in wireless communications.

**Yongxu Zhu** is a Professor in the National Key Laboratory of Mobile Communications, Southeast University (SEU). Her research interest is wireless communications theory.

**Kai-Kit Wong** is a Professor in the Department of Electronic and Electrical Engineering, University College London (UCL). His research interest is wireless communications theory.

**Kin-Fai Tong** is a Professor in the Department of Electronic and Electrical Engineering, University College London (UCL). His research interests are antenna theory and wireless communications.

**Hyundong Shin** is a Professor in the Department of Electronic Engineering, Kyung Hee University. His research interest is wireless communications theory.

**Tie Jun Cui** is a Professor in the State Key Laboratory of Millimeter Waves, Southeast University (SEU). He is an academician of Chinese Academy of Sciences. His research interests are electromagnetic theory, computational electromagnetics and wireless communications.